\documentclass[aps]{article}
\usepackage{geometry}              
\usepackage{graphicx}
\usepackage{subfig}
\usepackage{amssymb}
\usepackage{epstopdf}
\usepackage{amsmath}
\usepackage{setspace}
\usepackage{amsfonts}
\usepackage{bm}
\usepackage{hyperref}

\begin{document}
\bibliographystyle{prsty}

\title{Persistent Thermal Inhomogeneities in a Gas-Cluster Mixture}  
\author{Clifford Chafin\\\
\small{Department of Physics, North Carolina State University, Raleigh, NC 27695} \thanks{cechafin@ncsu.edu}}

\date{\today}
\maketitle

\begin{abstract}
Surface tension of small grains and droplets makes them stable only at a much lower temperature than in bulk.  This makes spontaneous nucleation unfavorable in many cases.  Kinetic approaches are delicate in that one can easily generate models that do not agree with thermodynamics in the large N limit.  
Here it is shown that thermodynamics itself dictates a kind of temperature suppression inside each small cluster in any gas-cluster mixture.  This gives a different perspective on the ``translation-rotation'' paradox in that this gives a time averaged steady state thermal inhomogeneity rather than just temporal fluctuations in the energy.  
This not only reduces the barrier to nucleation but also suggests a change in the thermal radiation spectrum from such a mixture that is not just a result of the inhibited radiation spectrum from Mie radiators.  Either verification or refutation of this effect will be shown have important consequences for thermodynamics.  An understanding of this effect will be essential to kinetic approaches to nucleation theory.  
\end{abstract}


\section{INTRODUCTION}
The formation of dust and droplets occurs by a much studied and yet still unclear process of nucleation.  The stability of such particles is always problematic because the surface tension (equivalently surface energy) of such a small body grows as a fraction of its total energy as the diameter decreases.  This makes small clusters less thermodynamically stable than large ones at the same temperature.  The induced pressures on nanometer sized droplets can be hundreds of atmospheres.  To reach the conditions where few body clusters are stable we need to be at temperatures far below the bulk boiling point.  Once droplets above a critical radius have formed, classical nucleation theory (CNT) dictates such clusters steadily grow from there to form droplets in a process that is only halted by a new equilibrium of vapor concentration and temperature.  A big problem is that often these predictions give growth rates far from what is observed.  Seeding or large fluctuations are necessary to overcome the barrier of the critical radius \cite{Kash00}.  Kinetic approaches are appealing but one must be careful to get enough details right so that thermodynamics actually emerges in the macroscopic limit.  

An interesting side note in the history of CNT is the introduction of translational and rotational energies of the embryonic clusters into the partition functions for the free energy by Pound and Lothe \cite{Lothe}.  This gave an enormous error in the predicted nucleation rates from theory and experiment by $\sim10^{17}$.  The subject seems to have closed with the work of Reiss, Kegel and Katz \cite{Reiss} which sought to explain it away through a discussion of length scale effects on entropy.  Although this is not going to be a paper on nucleation theory, we mention this history because it represents the dominant line of thinking of the interaction of clusters with an ambient gas.  

I will demonstrate here that, in the limit of very moderately sized clusters, the thermodynamic approach itself is somewhat inadequate and that the effect of these net cluster motions is to create net inhomogeneities in temperature. Specifically, the clusters will be  cooler than the ambient environment (i.e.\ the surrounding gas).
These will be smaller than the typical fluctuations but long lasting enough to make the temperature well defined for the purposes of droplet growth and thermal radiation.  The implicit assumption here is that most collisions of the clusters with the gas are thermalizing in that they transfer energy and momentum between the gas atoms and the internal and translational motion of the cluster rather than form binding or evaporation events.  

This seems initially paradoxical.  We are used to systems having a well defined global temperature for large $N$ and there being transient local fluctuations in energy, pressure, etc.  Temperature itself is typically defined through the microcanonical ensemble in terms of the density of states about a given net energy.  The introduction of temperature gradients requires a kind of ad hoc partitioning of the system in hopefully meaningful subparcels.  Here will will argue that such partitioning is not meaningful for this system.  The role of inhomogeneity of the gas-cluster system is to allow large enough clusters so that they can have very well defined temperatures internally but not so large that they have translational and rotational energy that can be neglected.  This observation sets it apart from the point of view of nucleation theories that rely on a precisely well-defined temperature in the $N\rightarrow\infty$ limit that achieve nucleation purely by local fluctuations in the free energy.  

\section{PERSISTENT INHOMOGENEITIES}
Let us redefine the problem as that of a set of uniform clusters that are unchanging in size and composition (so there is no active condensation or evaporation) in an ambient gas and investigate how the energy of the system is partitioned among them.  We ignore the role of the ambient radiation field, black body or external, in determining the distribution of energy.  This can be justified either by assuming that the system is essentially transparent to very cool surroundings and the radiative losses from the system are slow or that the transfer of heat among clusters is mediated by the gas at a much faster rate than the radiative field.  

If clusters are small and cold enough they cannot contain internal thermal excitations.  To apply equipartition the system must be in a regime far above this.  Treating the particles as solid cubes bound by SHO potentials of frequency $\omega_{0}$, we can specify the temperature at which this will be valid.  The longitudinal phonons with lowest energy have speed $\omega/k=v_{s}$.   We require $kT>>\hbar \omega(k)=h v_{s}/L$ or  $L>>h v_{s}/kT$.  The speed of sound in many forms of condensed matter is $v_{s}\sim 10^{3}$~m/s so, at room temperature, $L>>2\AA$.  For clusters of several nanometers across the approximation will typically be quite good.  

The equipartition theorem gives the mean energy per particle in the regime where quantum degeneracy is negligible.  Consider a set of freely floating clusters of $N$ particles in a gas.  These clusters have been assumed to be large enough at the ambient temperature for the internal energy states to form a quasicontinuum and that each modes is multiply occupied.  This justifies a classical treatment of their thermodynamics and assigning each cluster a well-defined temperature.   
The translational energy per particle is $\frac{3}{2}kT$.  This is divided among the internal motions of the cluster and the cm motion of it.  Since the cluster motion is independent of the motion of the internal particles we can partition the kinetic energy as $E_{\text{cm}}+E_{\text{internal}}$.  
This means that the mean effective temperature of the cluster (independent of its motion) is $T'=\frac{N-1}{N}T$.  It is this temperature which determines its stability and radiative properties.  For large clusters this is negligible but for temperatures around 1000~K we see that $N=100$ gives a 100~K temperature difference.  This is a significant effective cooling of the clusters and favors smaller cluster stability.  If the system is at an energy where rotational modes for the cluster are possible we have $T'=\frac{N-\frac{5}{3}}{N}T$ \footnote{In general, both modes will be possible at any relevant temperature but there is an interesting example where we expect it not to be: condensing $^{4}$He below the \protect $\lambda$-point.  Small droplets would tend to exclude vortices so would have no intrinsic angular momentum.  Furthermore, any angular momentum in the condensing gas would have to stay in the bulk motion and not impart any of it to the rotation of the condensing droplets.}.  

We generally expect a fluctuation in the temperature of these particles on the order of $T/\sqrt{N}$ due to finite size effects.  The temporal rate of these is not fixed by thermodynamics but will depend on the density and mass of the gas particles.  These two competing effects  leaves the mean temperature of the cluster unchanged but this preliminary observation suggests the thermal fluctuations will be larger than the shift by a factor of $\sqrt{N}$.  We will shortly examine this as well as the thermal equilibration times for the clusters.  

This ``translational depletion'' of the internal energy gives the depressed cluster temperature.  For completeness we should investigate why the presence of the moving spheres do not alter the gas distribution significantly.  Thermal velocities in the gas are $\sqrt{m_{c}/m_{g}}$ greater than the cluster.  Since the densities most solids and that of individual gas atoms are generally within an order of magnitude we have the velocity ratio  $v_{c}/v_{g}\approx N^{-3/2}$ and the cross sections of the clusters and gas related by $\sigma_{c}\approx N^{2/3}\sigma_{g}$.  

If the mean separation of clusters is much greater than the mean free path of the gas, $\lambda_{g}$, we can estimate the collision frequency of gas-cluster collisions, $\nu_{gc}$, from the number of gas molecules in a $\lambda_{g}^{3}$ volume about each cluster and the thermal velocity:  
\begin{align}
\nu_{gc}\sim \frac{n_{g}\lambda_{g}^{3}}{\tau_{g}}\frac{\sigma_{c}}{\lambda_{g}^{2}}=n_{g}\sigma_{c}v_{th}.
\end{align}
The frequency of the gas-gas collisions is $\nu_{gg}\approx v_{th}/\lambda\approx v_{th}n_{g}\sigma_{g}$.  The ratio of these collision rates is $\nu_{gc}/\nu_{gg}\sim N^{2/3}$.  This greater collision rate of the gas with the clusters is easily offset by the much greater number of gas particles, $n_{g}>>n_{c}$, and the fact that the greater mass and slower speed of the clusters makes gas-cluster collisions have a much smaller effect on the gas equilibration that gas-gas collisions.  

The duration that a given temperature persists for in each cluster depends on the fluctuations imparted by the gas and the rate at which it can thermally equilibrate and radiate and absorb energy.  If the cluster is small the radiative rate is reduced both by Raleigh/Mie effects and the reduced opacity from dimensions that are still smaller than the skin depth of the medium.  Since radiative cooling tends to be much less than from molecular collisions let us estimate equilibration time for our cluster.  This can be found from the specific heat, thermal conductivity, and size of the cluster $L$.  Specifically, $\Delta t\sim \frac{N_{c} k_{B}}{\kappa_{T}L}\approx \frac{\eta_{c} k_{B}}{\kappa_{T}}\sigma_{c}$ where $\eta_{c}$ is its number density.  

Any reasonable numbers for these parameters gives an astronomically much shorter equilibration time than the time between gas collisions with the cluster $\nu_{gc}^{-1}$.  The heating fluctuations must come from a long sequence of collisions that are biased above or below the mean thermal velocity of the gas and the cluster has a well defined temperature during most of the interim.  Each such collision has the bias in the temperature imparted reduced by $N^{-1}$ so thermal fluctuations in the gas on the scale of $\Delta T$ are moderated to $\Delta T/N$ in the cluster.  From this we can see that the clusters have well defined internal temperature for very much longer than characteristic times of the gas.  Furthermore, it suggests that our previous estimate of energy fluctuation $\sim\sqrt{N}$ in the cluster may need to be reduced by another factor of $N$ in a mixture that has its thermal properties dominated by the gas.  

\section{THERMAL RADIATION SPECTRUM}

The well defined thermodynamics of the cluster are what governs it's radiational heat losses and equilibration and these are determined entirely by the internal motions of the body.  These effectively colder clusters reduce the barrier to their formation by reducing evaporative losses.  Of course, nucleation is a gradual process but the above argument shows that the clusters that exist will equilibrate by gas collisions to give the usual thermal energy $\frac{3}{2}kT$ (or $\frac{5}{2}kT$ allowing bulk rotations).  This means there will be a deficit of internal energy in the internal motions.  We know that particles fixed at rest at temperature $T$ give the corresponding gray body radiation associated with that temperature.  Depressing this temperature means a cooler contribution to this field.  

Many nucleation environments are dilute, transparent and confined enough that this self generated field is not the source of the ambient radiation field.  However, in a highly reflective chamber, when the extent of the mixture is large or the clusters are comparable the the thermal wavelength and highly opaque, this will determine the radiative distribution emanating from the cloud.  

Smaller clusters where the internal thermal modes have large enough energy gaps that they are not sufficiently occupied introduce their own problems.  Let us assume we have large enough clusters to ignore this.  Raleigh/Mie size effects attenuate and shift the thermal radiative spectrum from the clusters \cite{Weldon}.  This is a difficult effect to theoretically predict.  Details of how the Planck black body spectrum arises has been elusive since it was derived.  In a black body cavity, the variations in emissivity with frequency and non-Lambertian reflection properties get washed out by an increased number of reflections in those modes.  

Opaque solids can be thought of as having internal black body fields that then escape at the surface.  Bodies that are small on the order of thermal radiation will tend to exclude those modes.  When they are partially transparent, the internal radiation field cannot equilibrate and create an internal black body-like spectrum since the energy exchange of the electromagnetic field to the phonons is of short duration (``The invisible man is blind - but warm'').  In lieu of a theoretic calculation for comparison we can appeal to experimental data.  
One can observe the spectrum from similar clusters that are fixed and isolated or embedded in a transparent medium (with $\epsilon\approx 1$).  This will specify a radiational spectrum for each temperature.  

We shall now design a cavity suitable to measure the predicted temperature suppression in the clusters.  
Let the mixture be placed in a highly reflective half transparent container small enough extent and at densities where the gas is essentially transparent.  Introduce clusters that are large enough to have high opacity to the range of frequencies near the radiative maximum and at a low enough density that their mean separation is much longer than maximum thermal wavelength of the clusters.  (We should be careful not to create a system that is many times deeper than the full optical depth.  This may tend to create a kind of opaque body that creates its own internally equilibrated radiation field that is hard to analyze.)  

No container is perfectly reflective, especially over a broad range of frequencies.  To ensure that the clusters dominate the spectrum choose a density that is optically opaque in this range.  This is equivalent to assuming that every path from a point on the edge into the cavity terminates on a cluster.  Diffraction may allow some bending around particles so this strictly gives a lower bound on the desired cavity size.  
 
This problem is reminiscent of the ``dark night sky'' or Olbers' paradox \cite{Olber}.  The clusters are separated by mean distance of $d=n_{c}^{-1/3}$.   Each successive shell at integral distances $j*d$ contain a quadratically increasing number of clusters that have a quadratically decreasing angular size.  This gives near opacity at distance $D=(n_{c} \sigma_{c})^{-1}$.  Given the cavity and the gas-cluster mixture we can now examine the radiative spectrum.  The arguments of the previous section imply a shift in the radiative spectrum by $\Delta T=T-T'$.  Since these clusters can be chosen small enough to make $\Delta T$ quite large this should be easily observed.  

We have implicitly assumed that the gas and clusters equilibrate and do not separate over the duration of the experiment.  The terminal velocity of cloud droplets is very slow and similarly we can choose a large enough gas density to ensure this true for the given clusters.

\section{CONCLUSION}

Nucleation theory has been dominated by thermodynamic models \cite{Doring} and these have had some numerical success \cite{Wolk}.  The corrections to these models are thought to reside somewhere in a kinetic theory approach but these models are often wildly off the mark.  Aside from the general question of what self consistency conditions are required for such a model to give thermodynamics in homogeneous large $N$ systems, an accurate understanding of the internal dynamics of the clusters, thermodynamic or molecular dynamic, will be necessary.  It is to this end that this work should be relevant to such approaches.  

The dynamic process whereby thermal radiation and evaporation occur is complicated and unclear so a kinetic approach to this process is difficult.  This seems necessary and unavoidable since the ratio of the various sizes of clusters and the correlation with their internal states  at a given temperature seems hard to determine thermodynamically.  Indeed, if such long lasting local variations in temperature are essential to the system it may be intrinsically impossible to solve the problem with thermodynamics.  The analysis here presumed a set of clusters large enough for equipartition to be valid.  In the regime of smaller growing clusters we will need a kind of kinetics that can capture the internal dynamics of the clusters and not just treat them as naive reservoirs.  As to how the radiative features change, this is an enduring problem and may have to wait for some new insight on the frontier of quantum statistical mechanics itself.  

A problematic feature in nonequilibrium thermodynamics is that temperature in statistical mechanics is intrinsically defined as a nonlocal feature.  For systems that are sufficiently homogeneous and slowly varying this seems manageable.  In the gas-cluster system the clusters can be large enough to locally equilibrate to have a well defined temperature on their own.  One could argue that we should locally redefine $T$ to the value in the gas at each cluster and avoid this problem.  However, this solution violates the maximizing of entropy of the overall system which ensures equipartition at high temperatures.  
If this is true, we have something possibly even more interesting that affects our notions of what thermal equilibrium really means.  

We have always assumed that maximizing entropy implies a constant temperature.  This exactly holds only in the limit of $N\rightarrow\infty$ but has previously held well for $N$ not very large at all.  The failure to do so here for arbitrarily large times,  large total $N$ and that persists for specific ``large'' subsets of the system seems new.  The existence of short time violations of the second law have been predicted and experimentally shown \cite{Jarz, Wang}.  This model may be a place to look for other thermodynamic surprises.

The author gratefully thanks North Carolina State University for support, Thomas Schafer for insightful discussions and encouragement and David Fallest for an education into the history of nucleation theory.  

\bibliography{gc} 

\end{document}